\documentstyle[epsfig,rotating]{elsart}
\newcommand{\lsim}{\stackrel{<}{\sim}}
\def \bfk{\mbox{\boldmath $k_\perp$}}
\begin{document}
%
%
\begin{frontmatter}
January 1997  \hfill  DESY 97--004  \\

\vspace*{0.5cm}


\vspace*{2.5cm}

\title{Summary of Physics Prospects for
Polarized Nucleon-Nucleon Scattering at HERA-$\vec N$ \thanksref{talk}
}
\thanks[talk]{Invited talk presented by W.--D.~Nowak at 
the 2nd ELFE Workshop (St. Malo, September 1996).}
\author[Zeuthen,Protvino]{V.A. Korotkov},
\author[Zeuthen]{W.-D. Nowak}
\address[Zeuthen]{DESY-IfH Zeuthen, D-15735 Zeuthen, Germany}
\address[Protvino]{IHEP, RU-142284 Protvino, Russia}
\begin{abstract}
The physics scope of polarized nucleon--nucleon collisions originating
from an internal polarized target in the HERA proton beam is
summarized.
Based on 240 pb$^{-1}$ integrated luminosity at 40 GeV c.m. energy,
statistical
sensitivities are given over a wide $(x_F, p_T)$--range for a variety
of inclusive and exclusive final states. By measuring
single spin asymmetries unique information can be obtained on
higher twist contributions and their $p_T$-dependence.
From double spin asymmetries in both photon and J/$\psi$ production
it appears possible to measure the
polarized gluon distribution in the range 0.1~$\leq x_{gluon}
\leq$~0.4 with a good statistical accuracy.
Drell-Yan pair production asymmetries in doubly longitudinal mode,
measurable in the mass range $3 \div 10$~GeV, allow discrimination
between different parametrizations of the polarized light sea quark
distributions. In doubly transverse mode access to the quark
transversity distribution might become possible.
Statistically significant results can be expected in the elastic
channel.

\end{abstract}
\end{frontmatter}

\newpage

%
\section{Introduction}
%
A complete theoretical picture of the nucleon structure will certainly
have to incorporate the polarization degree of freedom, both in its
perturbative and non-perturbative aspects. Although a largely
consistent
scheme exists since some time within the framework of perturbative
Quantum
Chromodynamics (pQCD) to describe short distance processes in strong
interactions, the long distance scales are far from being 
fully understood in QCD even when appreciating the considerable 
progress made in
recent lattice calculations (see e.g. \cite{goe1}). Spin measurements
have
always been sources of unexpected surprises; the most recent example
given by the simple and hitherto successful
Quark Parton Model which could not describe
the nucleon spin structure ('spin crisis') \cite{emc1}. This caused a
lot
of theoretical papers and the recent next-to-leading order
calculations \cite{Ball} were
found to reconcile the QCD improved Quark Parton Model with the
results of
the experiments. However, it is fair to say that
the present knowledge of the polarized gluon distribution is still
completely
insufficient and the same applies to the polarized sea. 
No measurements exist yet on polarized quark distributions
in case of transverse nucleon polarization. 

The `spin crisis' induced enormous theoretical and experimental
activities.
A number of new experiments was proposed to investigate in more detail
the
longitudinal spin structure of the nucleon by measuring double spin
asymmetries
in lepton nucleon collisions. Several measurements are already
completed (E142 \cite{142a} and E143 \cite{143a} at SLAC) and some are
still
taking and/or analysing data (SMC \cite{smc1} at CERN and E154
\cite{154a}
at SLAC). Others have just started (HERMES \cite{her1} at DESY) or
will
start data taking soon (E155 \cite{155a} at SLAC). Nevertheless, it
has to
be mentioned that purely inclusive measurements determining the
longitudinal
spin structure functions
g$_1$(x, Q$^2$) for proton, neutron, and deuteron are unfortunately
restricted to probe only certain combinations of the polarized
valence, sea,
and gluon contributions to the nucleon spin. A full analysis would
require additional inputs from other measurements to separate the
different
components. \\
With the target polarization vector being oriented perpendicularly to
the beam
direction the transverse spin structure of the nucleon becomes
partly accessible by
measuring the spin structure functions $g_2(x, Q^2)$ which contain a
twist--3 part that has recently been probed by SMC \cite{smcg2} and
E143
\cite{143g2}, although still with large error bars. We note here that
the
underlying QCD correlators are at the same time describing the
expected
performance of single spin asymmetries measurable in nucleon--nucleon
collisions \cite{sha1}.  \\
Semi--inclusive measurements with SMC \cite{smc2} and HERMES
\cite{her1} will
allow access to a variety of (combinations of) polarized parton
distributions
and to some polarized fragmentation functions. However, direct and
separate
measurements of the polarized gluon and sea quark distributions will
remain
of limited significance for quite some years.   

An experiment (`HERA--$\vec{N}$' \cite{desy96-095})
utilising an internal polarized nucleon target in the 820~GeV
HERA proton beam would constitute a natural extension of the studies
of the
nucleon spin structure in progress at DESY with the  HERMES
experiment \cite{her1}.
Conceivably, this would be the only place where to study high energy
nucleon--nucleon spin physics besides the dedicated RHIC spin program
at BNL
\cite{RSC} supposed to start early in the next decade. \\
An internal polarized nucleon target offering unique features such as
polarization above 80\% and no or small dilution, can be safely
operated
in a proton ring at high densities up to $10^{14}$ atoms/cm$^2$
\cite{ste1}.
As long as the polarized target is used in conjunction with an
unpolarized proton
beam, the physics scope of  HERA--$\vec{N}$ would be focused to
`phase~I', i.e. measurements of single spin asymmetries.
Once later polarized protons should become available, the same set-up
would
be readily available to measure a variety of double spin asymmetries.
These `phase~II' measurements would constitute an alternative
--~fixed target~-- approach to similar physics which will be
accessible to
the collider experiments  STAR and  PHENIX at the low end of
the RHIC energy scale ($\sqrt{s}~\simeq~50$~GeV) \cite{bun1}. 

We recall \cite{desy96-095} that the 
integrated luminosity calculation is based upon realistic figures. 
For the average beam and target polarisation
$P_B = 0.6$ and $P_T = 0.8$ are assumed, respectively. A combined trigger  and 
reconstruction efficiency of $C \simeq 50\%$ is anticipated.
Using $\bar{I}_B = 80 \; \mbox{mA} = 0.5 \cdot 10^{18} \; s^{-1}$
for the average HERA proton beam current (50\% of the design value)
and a rather conservative
polarized target density of $n_T = 3 \cdot 10^{13}$ atoms/cm$^2$ 
the projected integrated luminosity becomes
${\cal{L}} \cdot T = 240 \; pb^{-1}$
when for the total running time $T$ an equivalent 
of $T = 1.6 \cdot 10^7 \;s$ at 
100~\% efficiency is assumed. This corresponds to about 3 real years under
present HERA conditions. 
At present there are no arguments that the HERA design current of 160
mA should not be reachable in the polarized mode, as well.
In addition, 
experience from UA6 running at CERN shows that after having gained 
some practical running experience it presumably becomes
feasible to operate the polarized gas target at about 3 times 
higher density without seriously affecting the proton beam lifetime. 
Hence in a few years even 500 pb$^{-1}$ {\it per year} might presumably
become a realistic figure.
The sensitivities shown in the rest of the paper are calculated based
upon 240 pb$^{-1}$.

This paper intends to present a summary of the activities
undertaken so far to study the physics scope and the experimental
possibilities when placing an internal polarized target into the
(polarized)
HERA proton ring. In principle there exist three different future
options to
study polarized nucleon--nucleon interactions in fixed target mode at
HERA.
One obvious possibility would be to equip the future HERA West Hall
experiment
 HERA--B \cite{HERAb}
with a polarized internal target once the approved physics
program be concluded. The spectrometer, under assembly at present,
will have
a rather large acceptance,
a huge rate capability, and a quite complete particle identification
system.
Secondly, in the East Hall about 40\% of the floor is occupied
 by the HERMES
experiment whose rather limited aperture poses, however, considerable
constraints onto the nucleon--nucleon spin physics menu.
Nevertheless, rotating the whole set--up by $\pi$ and moving it into
the
proton beam line was kept as an option in the HERMES design from the
beginning,
although the high rate in the proton beam would require considerable
upgrades.
Last but not least, the remaining free space in the East Hall could be
used
for a completely new experiment in the proton beam line which could be
specifically
designed according to the spin physics requirements. However, at the
moment
any site discussion appears  premature and hence this study aims at
being as much as possible independent of a final site decision. 

\section{Single Spin Asymmetries}
%
Single spin asymmetries in large $p_T$ inclusive production, both
in proton-nucleon and lepton-nucleon interactions, have recently
received much attention (for references see \cite{desy96-04}).
The naive expectation that they should be zero in perturbative QCD has
been proven to be false, both experimentally and theoretically.
It is now clear that higher twist effects are responsible for
these asymmetries, which should be zero only in leading twist-2
perturbative QCD.  \\
Several models and theoretical analyses suggest possible higher
twist effects: there might be twist-3 dynamical contributions, which
we
shall denote as hard scattering higher twists; there might also be
intrinsic $k_\perp$ effects, both in the quark fragmentation process
and in the quark distribution functions. The latter are not by
themselves higher twist contributions - they are rather
non-perturbative
universal nucleon properties - but give rise to twist-3 contributions
when convoluted with the hard scattering cross sections.
The dynamical contributions result from a short distance part
calculable in perturbative QCD with slightly modified Feynman
rules, combined with a long distance part related to quark-gluon
correlations. 

An intrinsic $k_\perp$ effect in the quark fragmentation
is known as Collins or
sheared jet effect; it simply amounts to say that the number
of hadrons $h$ (say, pions) resulting from the fragmentation
of a transversely polarized quark, with longitudinal momentum
fraction $z$ and transverse momentum $\bfk$, depends on the
quark spin orientation. That is, one expects the {\it quark
fragmentation analysing power} $A_q(\bfk) \equiv
(D_{h/q^\uparrow}(z, \bfk) - D_{h/q^\downarrow}(z, \bfk)) /
(D_{h/q^\uparrow}(z, \bfk) + D_{h/q^\downarrow}(z, \bfk))
$ to be different from zero,
where, by parity invariance, the quark spin should be orthogonal
to the $q-h$ plane. Notice also that time reversal invariance
does not forbid such quantity to be $\not= 0$ because of the
(necessary) soft interactions of the fragmenting quark with
external strong fields, i.e. because of final state interactions.
This idea has been applied to the computation of the single
spin asymmetries observed in $pp^\uparrow \to \pi X$ \cite{art}. \\
A similar idea applies to the distribution functions, provided
soft gluon interactions between initial state partons are present and
taken into account, which most certainly is the case for hadron-hadron
interactions. That is, one can expect that the number of quarks with
longitudinal momentum fraction $x$ and transverse intrinsic motion
$\bfk$ depends on the transerve spin direction of the parent nucleon,
so that the {\it quark distribution analysing power}
$N_q(\bfk) \equiv 
(f_{q/N^\uparrow}(x, \bfk) - f_{q/N^\downarrow}(x, \bfk)) /
(f_{q/N^\uparrow}(x, \bfk) + f_{q/N^\downarrow}(x, \bfk))
$ can be different from zero.
This effect also has been used to explain the single
spin asymmetries observed in $pp^\uparrow \to \pi X$ \cite{ans}. \\
As mentioned above, both $A_q(\bfk)$ and $N_q(\bfk)$
are leading twist quantities which, when convoluted
with the elementary cross-sections and integrated over $\bfk$, give
twist-3 contributions to the single spin asymmetries. 

Each of the above mechanisms might be present and important
in understanding twist-3 contributions; in particular the quark
fragmentation or distribution analysing powers look like new
non-perturbative universal quantities, crucial in clarifying the
spin structure of nucleons. It is then of great importance to
study possible ways of disentangling these different contributions
in order to be able to assess the importance of each of them.
We propose here to measure the single spin asymmetry,
$ (d\sigma^{AB^\uparrow \to CX} - d\sigma^{AB^\downarrow \to CX}) \
  (d\sigma^{AB^\uparrow \to CX} + d\sigma^{AB^\downarrow \to CX}) $
in several different processes $A B^\uparrow \to C X$ which
should allow to fulfil such a task. To obtain a complete picture we
need
to consider nucleon-nucleon interactions together with other
processes,
like lepton-nucleon scattering, which might add valuable information.
For each of them we discuss the possible sources of higher twist
contributions, distinguishing, according to the above discussion,
between those originating from the hard scattering and those
originating
either from the quark fragmentation or distribution analysing power.

$\bullet \quad pN^\uparrow \to hX$:
all kinds of higher twist contributions may be present;
this asymmetry {\it alone} could not help in evaluating the relative
importance of the different terms.

$\bullet \quad pN^\uparrow \to \gamma X, \> pN^\uparrow \to \mu^+\mu^-
X,
\> pN^\uparrow \to jets + X$:
no fragmentation process is involved and we remain with possible
sources of non-zero single spin asymmetries in the hard scattering
or the quark distribution analysing power.

$\bullet \quad lN^\uparrow \to hX$:
a single spin asymmetry can originate either from
hard scattering or from $k_\perp$ effects in the fragmentation
function,
but not in the distribution functions, as soft initial state
interactions
are suppressed by powers of $\alpha_{em}$. 

$\bullet \quad
lN^\uparrow \to \gamma X, \> \gamma N^\uparrow \to \gamma X, \>
lN^\uparrow \to \mu^+\mu^- X, \> lN^\uparrow \to jets + X$:
a single spin asymmetry in any of these processes
may only be due to higher-twist hard scattering effects.

\newpage

It is clear from the above discussion that a careful and complete
study of single spin asymmetries in several processes might be a
unique way of understanding the origin and importance of higher twist
contributions in inclusive hadronic interactions; not only, but
it might also allow a determination of  fundamental non-perturbative
properties of quarks inside polarized nucleons and of polarized
quark fragmentations. Such properties should be of universal value
and applicability and their knowledge might be as important as the
knowledge of unpolarized distribution and fragmentation functions. \\
In the following we discuss the capability of HERA-$\vec N$ to
investigate single spin asymmetries. \\

\begin{figure}[ht]
\centering
\begin{minipage}[c]{6.8cm}
\raggedleft
\epsfig{file=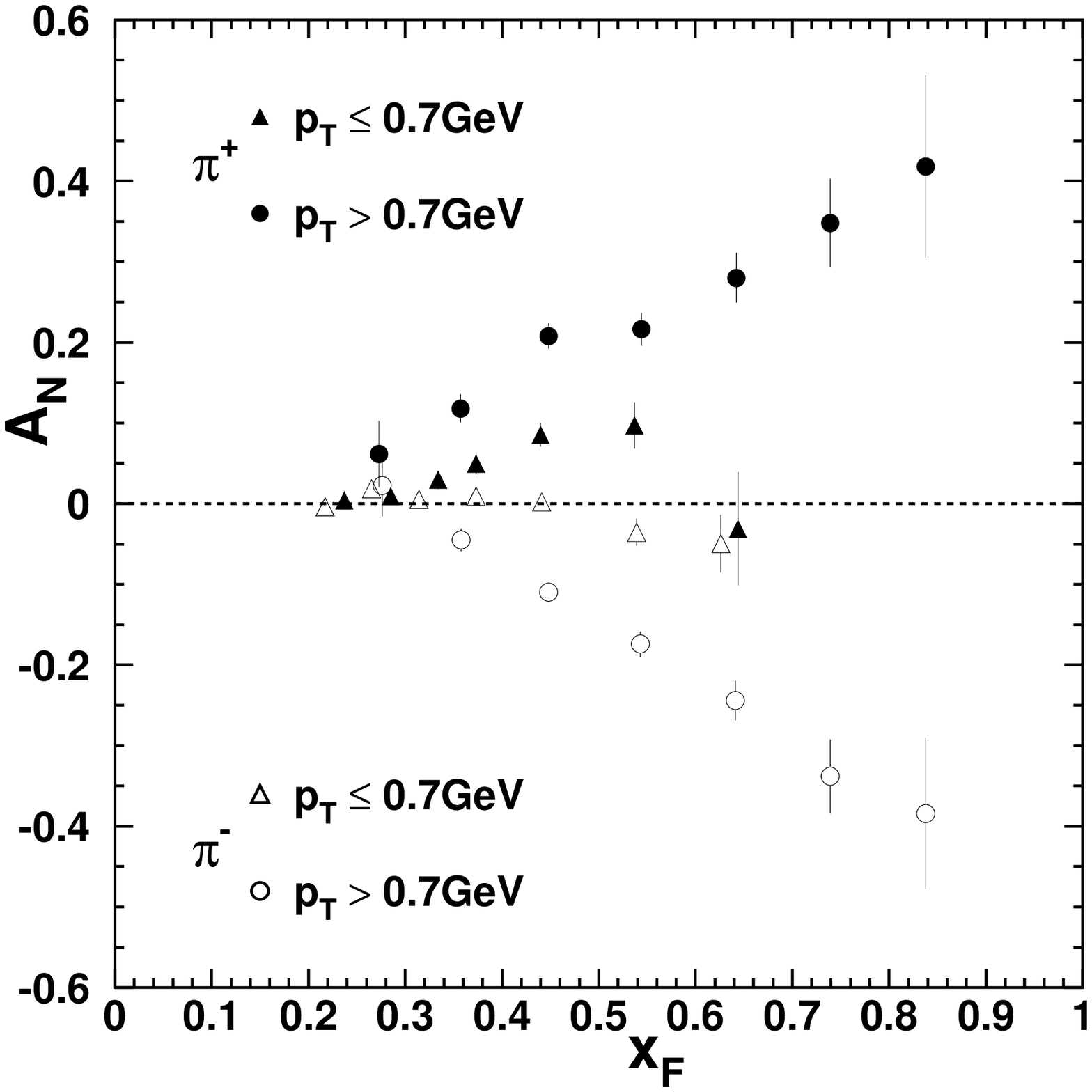, width=6.6cm}
\caption{\it Single spin asymmetry in inclusive pion production
         $p^{\uparrow}~+~p~\rightarrow~\pi^{0\pm}~+~X$ measured by
         the E704 Collaboration \protect \cite{704pi}
         and shown for two subregions of $p_T$. }
         \label{e704data}
\end{minipage}
\hspace*{0.2cm}
\begin{minipage}[c]{6.6cm}
\raggedright
\epsfig{file=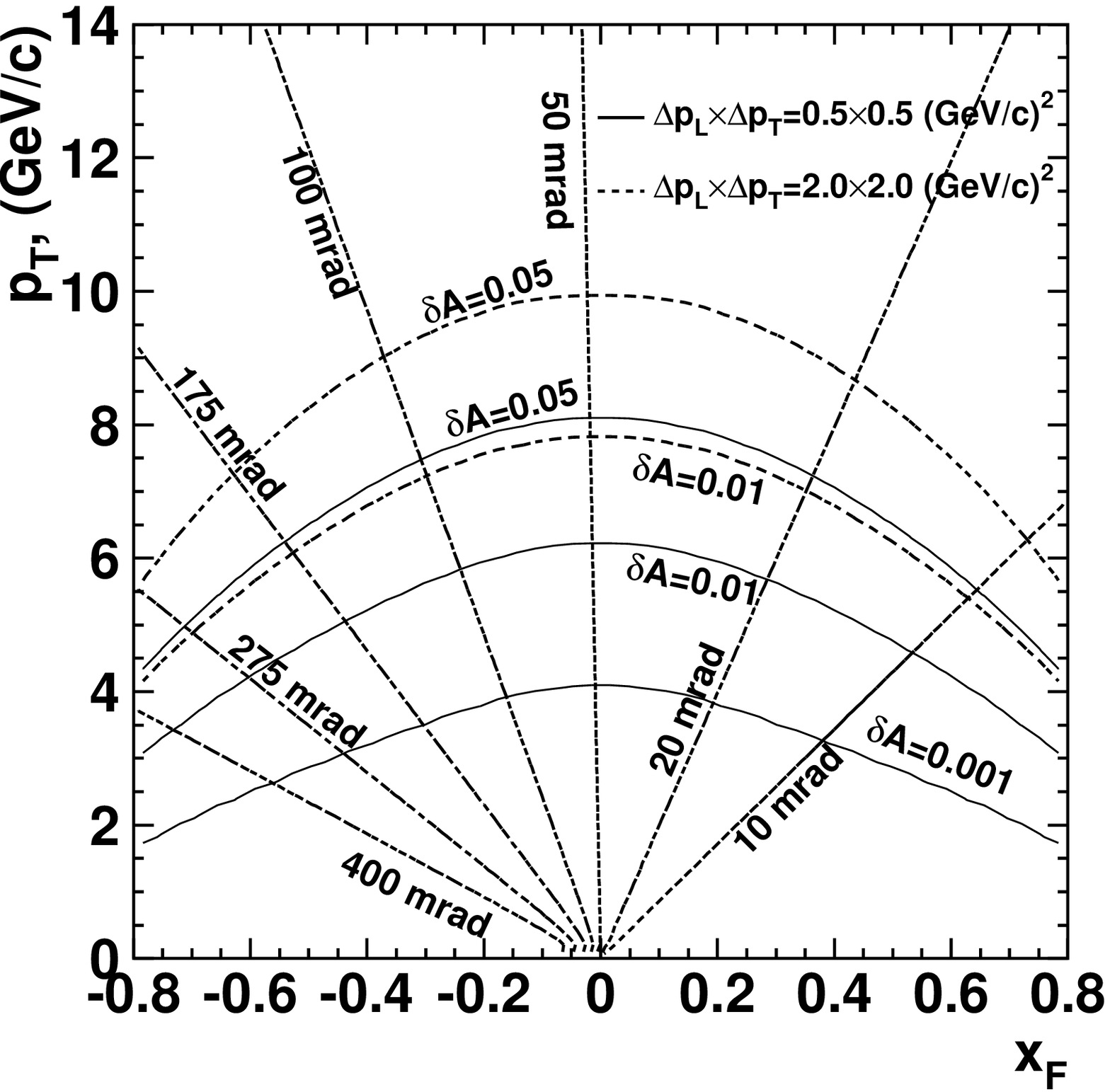,width=6.6cm}
\caption{\it Asymmetry sensitivity levels for $\pi^+$
       production in the ($p_T,~x_F$) plane. 
       Laboratory angles of the pions are shown.
        }     \label{fpisens}
\end{minipage}
\end{figure}

{\bf Inclusive pion production} $p^{\uparrow} p \rightarrow
\pi^{0\pm}X$
at 200 GeV exhibits surprisingly large single spin
asymmetries, as it was measured a few years ago
by the E704 Collaboration using a transversely
polarized beam \cite{704pi}. For any kind of pions the asymmetry $A_N$
shows a
considerable rise above
$x_F > 0.3$, i.e. in the fragmentation region of
the polarized nucleon. It is positive for both
$\pi^+$ and $\pi^0$ mesons, while it has the opposite sign for $\pi^-$
mesons.
The charged pion data taken in the $0.2 < p_T < 2$~GeV range
were split into two samples at $p_T$~=~0.7~ GeV/c; the observed rise
 is stronger for the high $p_T$ sample, as can be
seen from fig.~\ref{e704data}.

\begin{figure}[ht]
\centering
\begin{minipage}[c]{6.8cm}
\raggedleft
\epsfig{figure=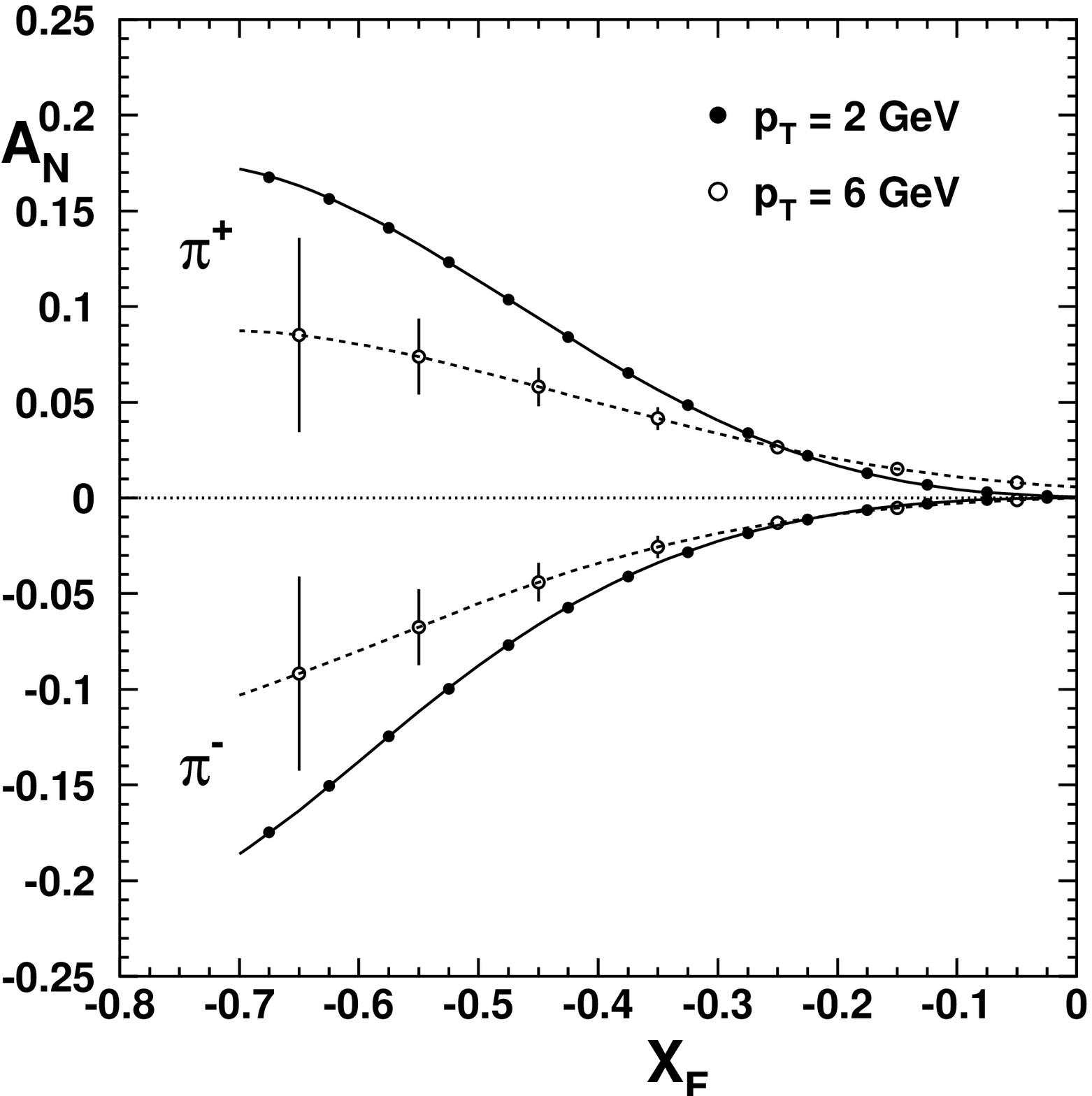,width=6.6cm}
\caption{\it Capability of HERA-$ \vec N $ to discriminate predictions
             for different $p_T$.   \protect\rule[-2\baselineskip]
 {0pt}{3\baselineskip}     }
             \label{asymurgia}
\end{minipage}
\hspace*{0.2cm}
\begin{minipage}[c]{6.6cm}
\raggedright
\epsfig{file=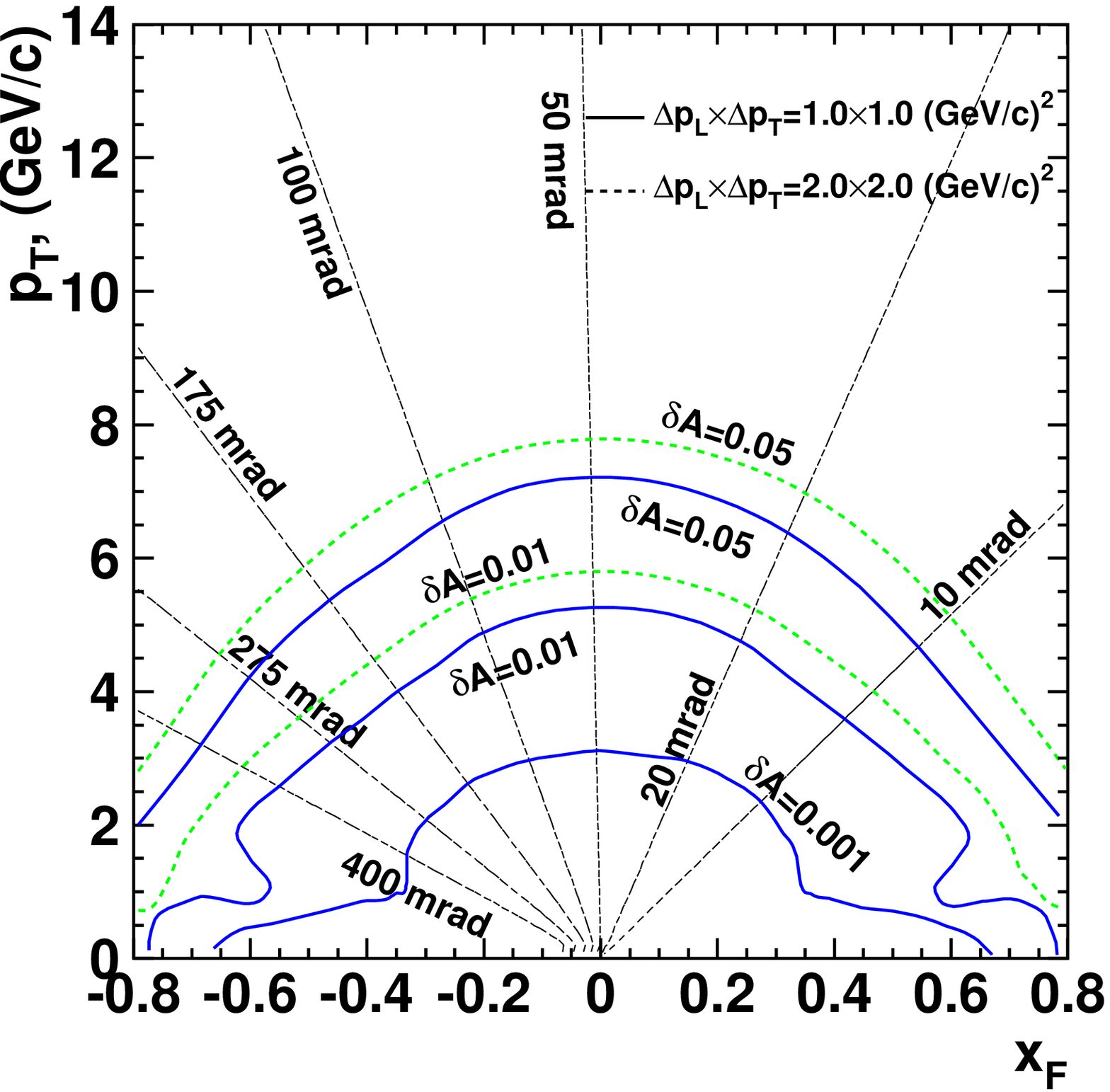,width=6.6cm}
\caption{\it Asymmetry sensitivity levels for photon
       production in the ($p_T,~x_F$) plane. 
       Laboratory angles of the photons are
       shown.
   }
     \label{gamxfpt}
\end{minipage}
\end{figure}

Contours characterizing different HERA-$\vec{N}$ sensitivity levels
($\delta A_N = 0.001$, $0.01$ and $0.05$) for an asymmetry
measurements
in the reaction  $pp^{\uparrow} \rightarrow \pi^+ X$ are shown in
fig.~\ref{fpisens}.
Note that in the large $p_T$ region the contours calculated with big
$\Delta p_L \times \Delta p_T$ bins are appropriate, since usually a
larger
bin size is chosen where the statistics  starts to decrease.
We can conclude that the accessible $p_T$ values are significantly
larger
than those E704 had; the combined $p_T$ dependence of all involved
higher-twist
 effects can be measured with good accuracy ($\delta A_N \leq 0.05$)
up to transverse momenta of about 10~GeV/c in the central region
$|x_F| < 0.2$
and up to 6~GeV/c in the target fragmentation region.
This corresponds to
an almost one order of magnitude extension in the $p_T$ range in
comparison
to E704. The capability of HERA-$\vec{N}$ to really prove a predicted
$p_T$
dependence is shown in fig.~\ref{asymurgia}, where the curve was
obtained
assuming a non--zero quark distribution analysing power,
$N_q(\bfk)$, according to Ref. \cite{ans}.  \\

The study of polarization asymmetries in {\bf vector meson production}
is especially attractive as these particles are produced 
`more directly' in comparison to pions which
are mainly decay products of heavier particles. 
Comparing asymmetries in vector and
pseudoscalar meson production can provide information on the 
magnitude of the asymmetry in quark scattering \cite{czyz}. 
If the asymmetry is generated only
during the fragmentation of polarized quarks~\cite{art}, 
the asymmetry of $\rho$ mesons is expected to be opposite 
in sign to that of pions,
$R_{\rho / \pi} = A_N^\rho / A_N^\pi \simeq -{1 \over 3}$.
On the contrary, if the quark scattering asymmetry  were 
the dominating one, the asymmetries of pseudoscalar and vector
mesons would not differ substantially. Note that 
$R_{\rho / \pi} \neq -{1 \over 3}$ could
also mean a violation of the non-relativistic quark model, which was
assumed in the calculation of the asymmetry. Although HERA-$\vec N$
specific projected errors remain to be calculated, significant
results can be expected over a wide kinematical region due to abundant
production of vector mesons.

\newpage

{\bf Inclusive direct photon production}, $p p^{\uparrow} \to \gamma
X$,
proceeds without fragmentation, i.e. the photon carries directly the
information from the hard scattering process. Hence this process
measures
a combination of initial $\bfk$ effects and hard scattering twist--3
processes. The first and only results up to now were obtained by E704
Collaboration \cite{Phot704}
showing an asymmetry compatible with zero within
large errors for $2.5 < p_T <3.1$~GeV/c in the central region
$ | x_F | \lsim 0.15$.   \\
The experimental sensitivity of HERA-$\vec N$ (see fig.~\ref{gamxfpt})
was determined using cross-section calculations
 of the two dominant hard subprocesses
contributing
to direct photon production, i.e. gluon--Compton scattering
($qg \rightarrow \gamma q$) and quark--antiquark annihilation
($q \bar q \rightarrow \gamma g$),
and of background photons that originate mainly from $\pi^0$ and
$\eta$ decays.
It turns out that a good sensitivity (about 0.05)
can be maintained up to $p_T \leq$ 8 GeV/c.
For increasing transverse momentum the annihilation subprocess and the
background photons are becoming less essential;
we expect to be able to detect a clear dependence on $p_T$,
of the direct photon single spin asymmetry.

\begin{figure}[ht]
\vspace*{-5mm}
\centering
\begin{minipage}[c]{6.8cm}
\vspace*{-3mm}
\raggedleft
\epsfig{figure=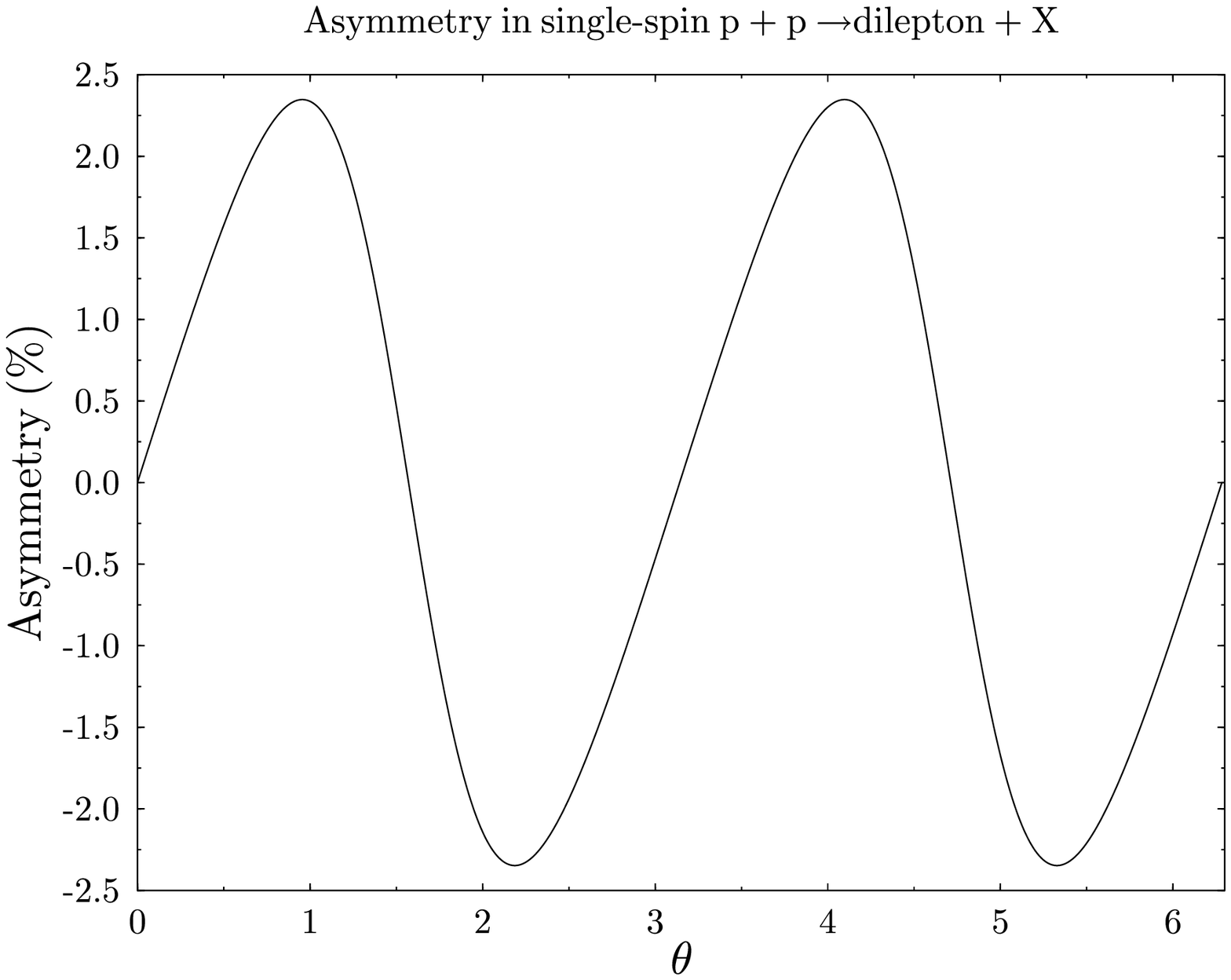,width=6.8cm}
\vspace*{1mm}
\caption{\it Single~spin~asymmetry 
  in Drell-Yan pair production
  at HERA-$\vec N$ energy, $x=0.5$ and $M^2=10~GeV^2$. }
  \label{ssady}
\end{minipage}
\hspace*{0.2cm}
\begin{minipage}[c]{6.6cm}
\raggedright
\epsfig{figure=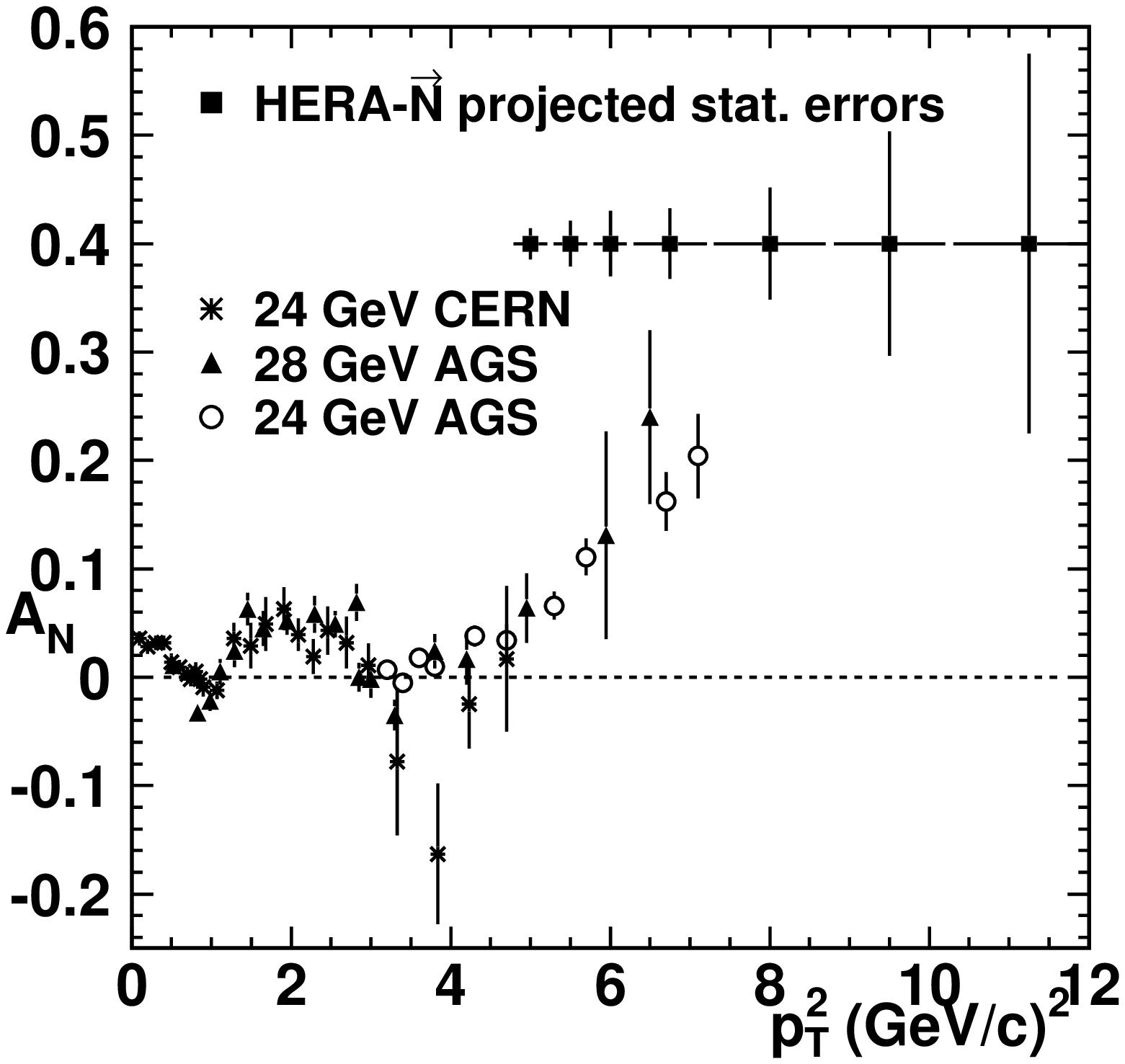,width=6.6cm}
\caption{\it Single spin asymmetry in polarized
  proton-proton elastic scattering as a function of $p^2_T$.
     \protect\rule[-2\baselineskip]
 {0pt}{3\baselineskip}  }
  \label{ppelastic}
\end{minipage}
\end{figure}

The single spin asymmetry in {\bf Drell-Yan production},
$p~+~p^{\uparrow}\rightarrow~ l \bar l~+~X$, at small
transverse momenta was calculated \cite{hammon} in the framework
of twist-3 pQCD at HERA-$\vec N$ energy. The resulting asymmetry
which does not exceed 2\%
depends strongly on the kinematical domain;
in fig.~\ref{ssady} it is shown
as a function of one of the lepton's polar angle for a particular 
kinematical situation, where
the momentum fraction of the scattered quark was chosen as $x=0.5$
and the  dilepton mass was fixed at $M^2 = 10$~GeV$^2$. 

{\bf Inclusive J$/\psi$ production} was 
calculated in the framework of the colour singlet model \cite{psicosi}.
The calculations at HERA-$\vec N$ energy \cite{desy96-04}
show an asymmetry less than 0.01 in the region $|x_F|~<~0.6$,
i.e. the effect is practically unobservable. 

Summarizing the prospects for the measurement of single spin 
asymmetries in inclusive reactions we conclude that an asymmetry
size of a few percent seems to be a `canonical' order of
magnitude for single spin asymmetries calculable in present 
twist-3 pQCD rather independent on the specific final 
state \cite{wdn-dubna-96}. 
Nevertheless,  one may expect larger 
predictions when combining the pure pQCD effects with non-zero transverse 
momenta in both the distribution and the fragmentation functions.
We note that asymmetries on the few percent level are
difficult to measure, even with sufficiently small statistical errors,
since the systematic error originating mainly from beam and target
polarization measurements constitutes a severe limit.

Large spin effects in {\bf proton-proton elastic scattering},
$p~+~p^{\uparrow}\rightarrow~p~+~p$, have been disco\-vered many years ago.
The single spin asymmetry
$A_N$ was found significantly different from zero as it is shown 
in fig.~\ref{ppelastic} in conjunction 
with the projected HERA-$\vec N$ statistical errors.  
At HERA-$\vec N$ energy
the detection of the recoil proton for 
$p^2_T$ values in the range $5 \div 12$ (GeV/c)$^2$ requires a very large 
angular acceptance (up to 40 degrees) \cite{desy96-04}. 
The forward protons 
for the same interval in $p^2_T$ have laboratory angles of the order
of a few milliradians and require a dedicated forward detector very 
close to the beam pipe. \\
The transverse single spin asymmetry $A_N$ in elastic $pp$ scattering 
at HERA-$\vec{N}$ and RHIC energies has been calculated in a
dynamical model that leads to spin-dependent pomeron couplings
\cite{golosk}. The predicted asymmetry is about 0.1 for 
$p_{T}^2~=~4~\div~5$~(GeV/c)$^2$
with a projected statistical error of $0.01~\div~0.02$ 
for HERA-$\vec{N}$ (cf. fig.~\ref{ppelastic}), i.e.
a significant measurement of the asymmetry $A_N$ can be performed
to test the spin dependence of elastic $pp$ scattering at high energies.

\section{Double Spin Asymmetries}

Perturbative QCD allows for a simple calculation of Born double spin
asymmetries for various $2 \rightarrow 2$ subprocesses at the partonic
level. The one-loop radiative corrections to various subprocesses
have now been calculated \cite{contog}, they produce only small
changes in the asymmetry in comparison with the leading order.
Relying on factorization a rich spectrum of hadronic level asymmetries
is predicted which constitutes the backbone of the RHIC spin physics 
program. \\
Still,
the insufficient knowledge of the polarized parton distributions makes the
predictions for double spin asymmetries 
to a large extent uncertain.
Conversely, the measurement of double spin asymmetries
in certain final states seems to be
among the most valuable tools to eventually determine
the polarized parton distribution
functions in the nucleon. The presently most accurate way to do so
is the study of those processes which can be calculated
in the framework of perturbative QCD, i.e. for which the involved
production cross sections and subprocess asymmetries can be predicted.
Production of {\it direct photon (plus jet)},
{\it J/$\psi$ (plus jet)} and {\it Drell-Yan pairs} are most
suited because there are only small uncertainties due to
fragmentation. \\
Once having available a polarized proton beam at HERA,
all possible combinations of beam and target polarization
($LL,~TT,~LT$) could be investigated at HERA-$\vec N$. This is especially
interesting since it would open access to polarized
parton distributions which cannot be studied in inclusive lepton
DIS. \\
In the following we discuss the capabilities of
HERA-$\vec{N} $, operated in doubly polarized mode (`Phase~II'),
to perform such measurements.

\begin{figure}[hbt]
\centering
\epsfig{figure=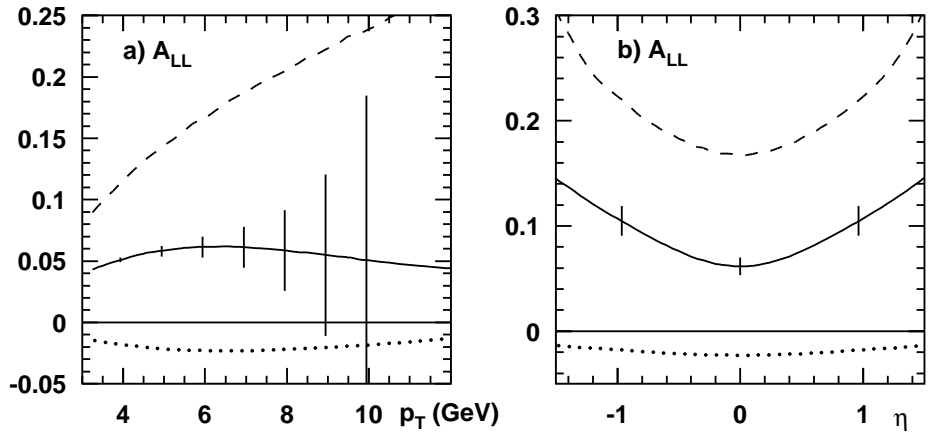,width=14cm}
\caption{\it Double spin asymmetry in inclusive photon production
displayed vs. a) $p_T$ and b) $\eta$ for different polarized gluon distribution
functions (see Ref. \cite{gor1}), shown in conjunction with the
projected statistical sensitivity of HERA-$\vec{N}$.}
 \label{gamvogel}
\end{figure}

{\bf Direct photon production} in $pp$~interactions is dominated 
by the quark-gluon Compton subprocess, $q(x_1) + g(x_2) \rightarrow
\gamma + q$. Ignoring the integration over the kinematical domain
the longitudinal double spin asymmetry can be written schematically in
the simple form
\begin{equation}
\label{gammall}
 A_{LL} = { {\sum_i e^2_i[\Delta q_i(x_1) \Delta G(x_2) + (1 
 \leftrightarrow 2)] } \over
   {\sum_i e^2_i[q_i(x_1) G(x_2) + (1 \leftrightarrow 2)] }}
  \hat a_{LL},
\end{equation}
where $\hat a_{LL}$ is the partonic asymmetry in the gluon Compton
subprocess. The hadron level asymmetry $A_{LL}$ is directly sensitive 
to the polarized gluon distribution. We note that in (any) 
{\it inclusive} production no direct relation can be established
between the measured asymmetry $A_{LL}$ and $\Delta G \over G$. \\
In a recent study \cite{gor1}  {\bf inclusive photon production}
with HERA-$\vec{N} $ was investigated. Basing on a NLO calculation
rather firm predictions were obtained for $A_{LL}$ including an
assessment of the theoretical uncertainties; the latter turned out
to be of rather moderate size. In fig.'s \ref{gamvogel}a and
\ref{gamvogel}b three different
predictions for the asymmetry are shown
in dependence on $p_T$ and pseudorapidity $\eta$, in conjunction with
the projected statistical uncertainty of HERA-$\vec{N} $.
As can be seen, there is sufficient
statistical accuracy in a wide kinematical region
to discriminate between different polarized gluon distribution
functions.

%
%
%
Compared to direct photon production the
production of $c \bar c$ quarkonium
states, in particular {\bf inclusive $J/\psi$ production},
 is a similarly clean tool to
measure the polarized gluon distribution.
For the production of quarkonia with $p_T$ above 1.5~GeV 
the $2 \rightarrow 2$
subprocess $g(x_1) + g(x_2) \rightarrow (c \bar c) + g$ provides the main
contribution. Consequently the asymmetry can be written as
\begin{equation}
\label{jpsill}
 A_{LL} = {{\Delta G(x_1)} \over {G(x_1)}} {{\Delta G(x_2)} \over
   {G(x_2)}} \hat a_{LL}.
\end{equation}
Because of the relatively large quark mass the $c\bar c$
production cross section and the expected asymmetry are supposed to be
calculable perturbatively.
The calculation of the longitudinal double spin asymmetry in
ref.~\cite{TT} takes into account both the color singlet and the
color octet states of the $(c \bar c)$-pair. The long distance matrix
elements responsible for evolving the color octet $(c \bar c)$
state into a quarkonium were extracted from experimental data.

\begin{figure}[ht]
\centering
\begin{minipage}[c]{6.8cm}
\centering
\epsfig{file=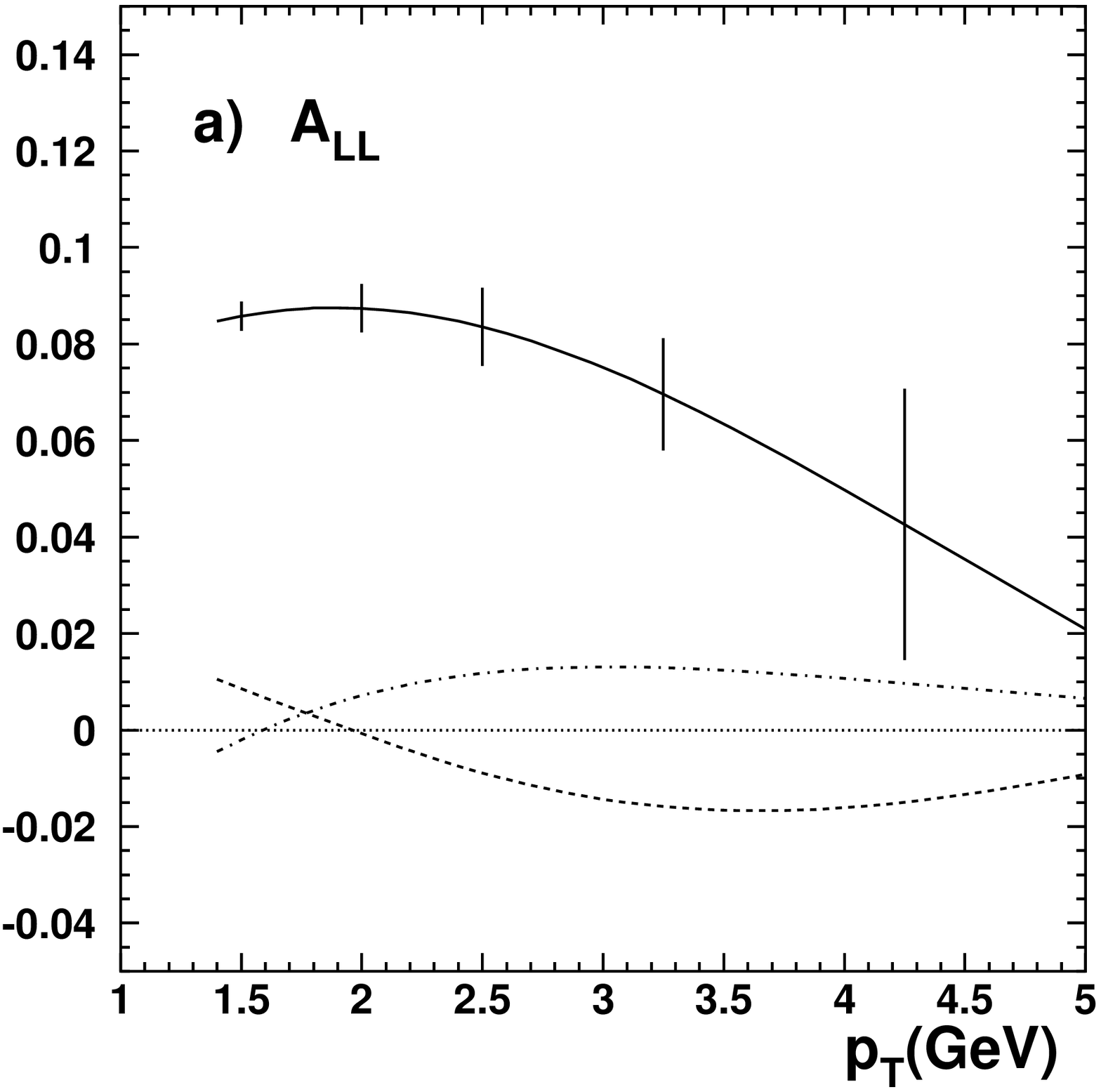, width=6.8cm}
\end{minipage}
\begin{minipage}[c]{6.8cm}
\centering
\epsfig{file=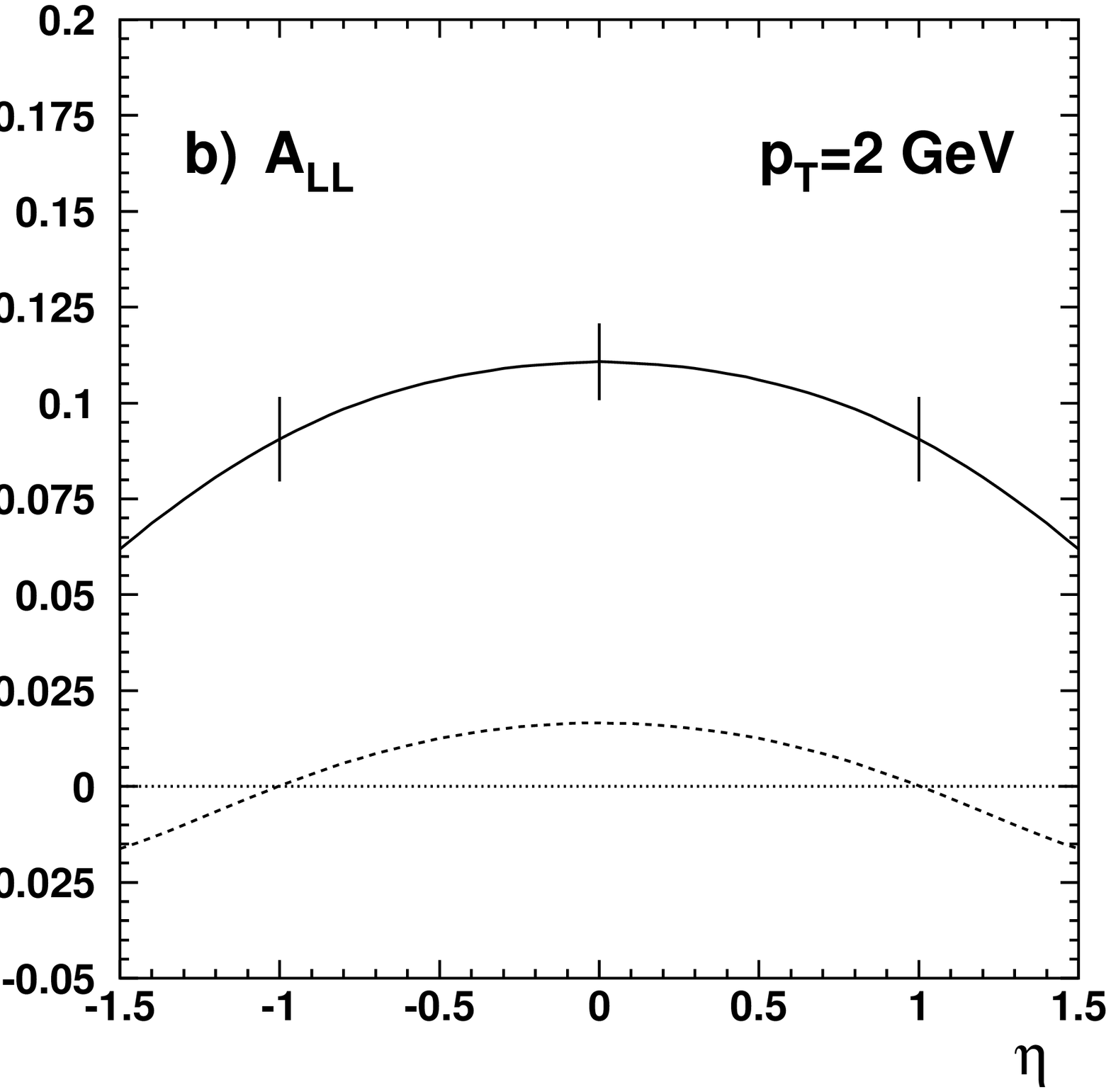, width=6.8cm}
\end{minipage}
\caption{\it Double spin asymmetry in inclusive $J/\psi$ production
displayed vs. a) $p_T$ and b) $\eta$ for 
LO  set~A (full line) and set~C (dashed line) of
Ref. \protect \cite{GehrStir}, shown in conjunction with the
projected statistical sensitivity of HERA-$\vec{N}$. 
The dash-dotted line in a)
represents the color singlet contribution to the asymmetry. }
\label{jpsiavto}
\end{figure}

In fig.'s \ref{jpsiavto}a and \ref{jpsiavto}b 
the expected asymmetry is presented versus
$p_T$ and $\eta$ for two different
polarized gluon distributions from ref.\cite{GehrStir};
apparently a very good discrimination
between different sets
is possible over the HERA-$\vec N$ kinematical range.
Fig.~\ref{jpsiavto}a shows also that  the color octet mechanism gives 
the main
contribution to the $J/\psi$ production asymmetry at HERA-$\vec N$
energy.

\begin{figure}[hb]
\centering
\epsfig{file=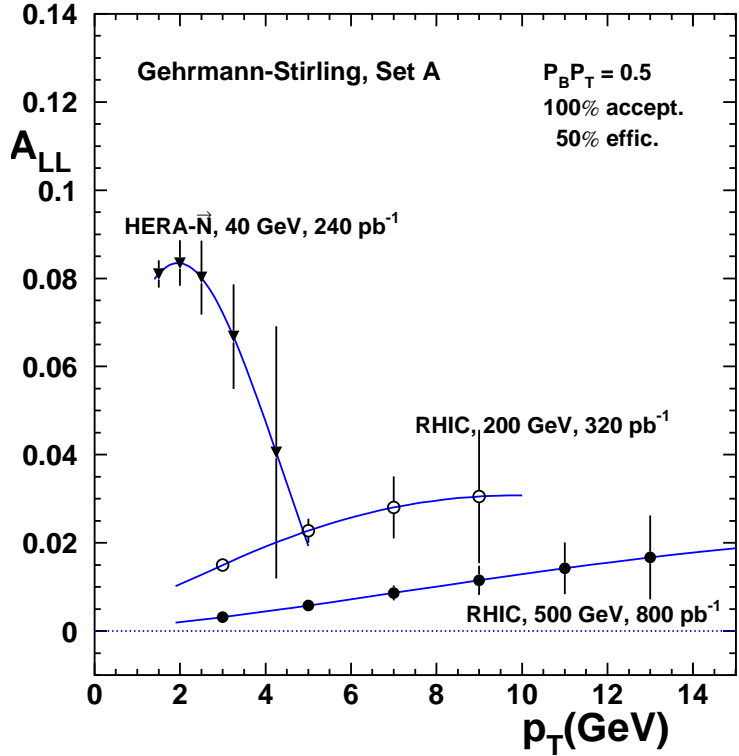, width=11.cm}
\caption{\it Transverse momentum
dependence of the expected asymmetries and projected
statistical errors for $J/\psi$ production
at HERA-$\vec N$ and for two different energies at RHIC. }
\label{jpsirhic}
\end{figure}

For  comparison the expected double spin
asymmetries for $J/\psi$   production at RHIC energies
were calculated \cite{KNT-Dubna}.
In fig.~\ref{jpsirhic} predictions at HERA-$\vec N$ and 
two different RHIC energies are shown in conjunction with 
the projected statistical errors calculated by integration of the 
differential cross sections over $p_T$
with bins of $\Delta p_T=0.5$ GeV.
In the statistically accessible $p_T$ interval the asymmetry
ranges between 0.08 and 0.04 at HERA-$\vec N$, but only
between 0.01 and 0.03 at RHIC energies.
Comparing both ranges
and taking into account the above mentioned limitations by
systematic errors it is likely
that the fixed target experiment might accomplish a more
significant measurement of the charmonium production asymmetry.

\hfill
 
\newpage

{\bf Photon or $J/\psi$ Production Associated with Jets}.
The complete kinematics of the
2$\rightarrow$2 subprocess can be reconstructed if the away-side jet
in the production of photon or $J/\psi$ is measured as well. In this
case the asymmetry $A_{LL}$ can be directly related to the polarized
gluon distribution if a certain subprocess can be selected (for more details
see Ref. \cite{desy96-04}). 
\\
The measurement of double spin asymmetries in {\it photon (or J/$\psi$)
plus jet} production requires rather similar scattering angles for both
emerging `partons' in the laboratory system. Especially the jet must not
escape under too small angles, otherwise it can not be detected
efficiently under the given fixed target conditions (for more details
see Ref. \cite{desy96-04}). Hence a rather forward oriented detector with
good granularity down to scattering angles of $10 \div 20$ mrad is required.

\begin{figure}[hbt]
\centering
\epsfig{file=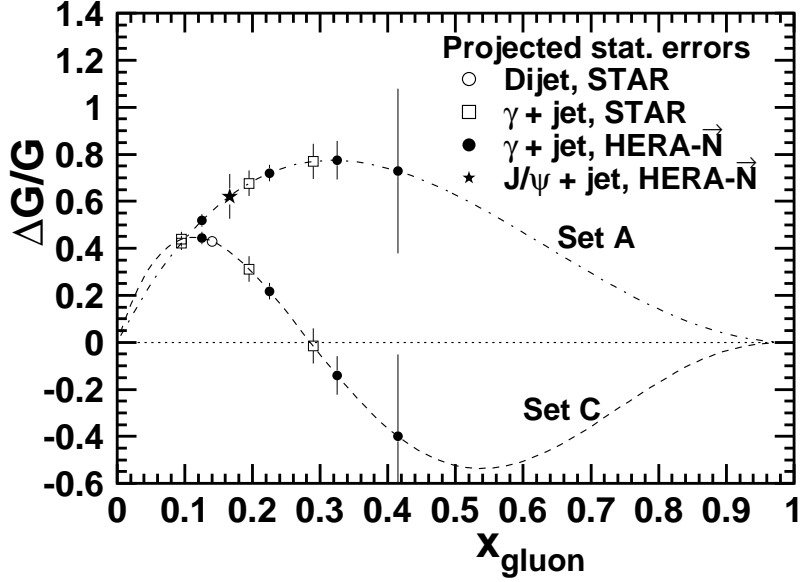,width=12cm}
\caption{\it Typical predictions for the polarized gluon
       distribution (LO calculations
       from Ref. \protect \cite{GehrStir}) confronted
       to the projected statistical errors expected for
       HERA-$\vec{N}$ and RHIC experiments.  }
      \label{gamjet}
\end{figure}

Using this approach {\bf photon plus jet} production was discussed in
Ref. \cite{96-128} as a tool to directly measure $\Delta$G/G.
The quark--antiquark annihilation subprocess is suppressed relatively
to quark-gluon Compton scattering because of the lower density of
antiquarks
(of the polarized sea) compared to gluons (polarized gluons).
In fig.~\ref{gamjet} the projected statistical sensitivity of 
HERA-$\vec{N}$ for the
$\Delta G(x)/G(x)$ measurement,
on the present level of understanding, is shown vs. $x_{gluon}$ in
conjunction
with predicted errors for  STAR running at RHIC at 200 GeV
c.m. energy \cite{yok1}.
The errors demonstrate clearly that in the region
$0.1 \leq x_{g} \leq 0.4$ a significant result from {\it photon plus
jet production} can be expected from HERA-$\vec N$
 with an accuracy being about competitive
to that predicted for RHIC.

In {\bf $J/\psi$ plus jet} production the quark-gluon subprocess 
contributes only about 10\% to the asymmetry compared to the 
gluon-gluon fusion subprocess.
Here the measurement of  $\Delta G(x)/G(x)$ is
feasible only for $x_{gluon}~=~0.1~\div~0.2$, i.e. for $J/\psi$ transverse
momenta of about 2.5 GeV/c, being a perturbatively safe value in
quarkonium production. This prediction is shown as an
additional entry in fig.~\ref{gamjet}. Although being one point
only, this is an important measurement because
the lowest lying point from {\it photon plus jet} production at
HERA-$\vec N$ is obtained
for rather small values of $p_T$ where pQCD is not
expected to give reliable predictions in this channel.
We note that the nature of the
gluon-gluon subprocess has similar consequences for
{\it dijet } production at RHIC; there is \cite{yok1} only
one point at a similar value of
$x_{gluon}$, as well (cf. fig.~\ref{gamjet}). \\
At this point we note that
the HERA-$\vec{N} $ fixed target kinematics causes additional problems
for the jet reconstruction when compared to a collider experiment.
The number of photon events accompanied by a successfully
reconstructed jet
decreases considerably when approaching
lower values of $p_T$ and, correspondingly, of $x_{gluon}$.
Preliminary jet reconstruction efficiencies were calculated and taken
into account for the results in fig.~\ref{gamjet}. \\
Although the $x_{gluon}$ interval ($0.1 \div 0.4$) explored by both
HERA-$\vec N$ and RHIC is quite comparable, the different 
transverse momentum ranges accessed (2...8 GeV at HERA-$\vec N$;
10...40 GeV at RHIC) make both measurements indeed complementary;
${\Delta G} \over G$ would be studied by HERA-$\vec N$ in the pQCD 
onset region whereas the RHIC experiments will explore 
${\Delta G} \over G$ in the deep perturbative region.

\begin{figure}[hbt]
\centering
\begin{turn}{-90}
\epsfig{file=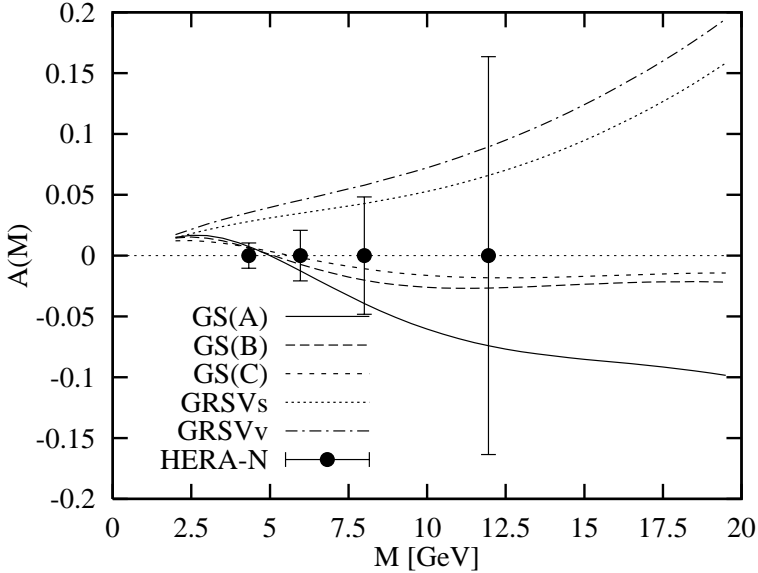,width=8cm}
\end{turn}
\caption{\it Expected asymmetries in the polarized Drell-Yan process
       from Ref. \protect \cite{DYGehrStir}) confronted
       to the projected statistical errors expected for
       HERA-$\vec{N}$.  }
      \label{dygehr}
\end{figure}

The production of {\bf Drell-Yan pairs} in nucleon-nucleon 
collisions proceeds via
quark-antiquark annihilation into a massive off-shell photon  which
subsequently decays into a lepton pair, 
$q(x_1) + \bar q (x_2) \rightarrow \gamma ^* \rightarrow l^+ l^-$.
The {\bf longitudinal} double spin asymmetry
turns out to be well suited to extract the
polarized light sea-quark distribution, since schematically
\begin{equation}
\label{DYLL}
A^{DY}_{LL} = {{\sum_i e^2_i [\Delta q_i(x_1) \Delta \bar q_i(x_2) + (1
    \leftrightarrow 2)]} \over 
{\sum_i e^2_i [q_i(x_1) \bar q_i(x_2) + (1
  \leftrightarrow 2)]} } \hat a_{LL}
\end{equation}
with $\hat a_{LL} = -1$. An experimental study of this asymmetry
can be performed with the HERA-$\vec N$ experiment; the prospects
for such a measurement were calculated  in ref.~\cite{DYGehrStir} 
at next-to-leading order QCD. The asymmetries expected  for 
different parametrizations of the polarized
NLO parton distributions are shown in
fig.~\ref{dygehr} in conjunction with the statistical accuracy
achievable at HERA-$\vec N$. The spread of the predictions reflects
the insufficient present know\-ledge on the polarized sea quark
distributions in the region $x > 0.1$; not even the sign of the
asymmetry at large $M$ is predicted. Since the asymmetry is the weighted 
sum of $\Delta \bar u$ and $\Delta \bar d$ quarks with the strange
quark contribution assumed to be small and the weight of $\Delta \bar u$ 
is higher than that of $\Delta \bar d$ due to its abundance in the
proton and the electric charge, the measured asymmetry provides mainly 
information on $\Delta \bar u$, i.e. on the light $u$ quark polarization.

{\bf Drell-Yan pair} production with {\bf transverse polarization} of
both beam and target can provide a qualitatively new insight
into the spin structure of the nucleon by measuring the third twist-2
quark distribution function (often called quark
transversity distribution, $\delta q(x)$ or $h_1(x)$) which is
absolutely unknown at the present time.
It basically describes the fraction of
transverse polarization of the proton carried by its quarks. 
In inclusive lepton DIS a contribution of this function is suppressed 
by a quark mass whereas it is in principle accessible in 
semi-inclusive DIS \cite{her1,JaffeJi1}.
The transverse double spin asymmetry in nucleon-nucleon
 Drell-Yan production 
can be schematically written in the form \cite{JaffeJi2}
\begin{equation}
\label{DYTT}
A^{DY}_{TT} = {{\sin^2 \theta \cos{2 \phi}} \over {1 + \cos^2 \theta}}
{{\sum_i e^2_i [\delta q_i(x_1) \delta \bar q_i(x_2) + (1
    \leftrightarrow 2)]} \over 
{\sum_i e^2_i [q_i(x_1) \bar q_i(x_2) + (1
  \leftrightarrow 2)]} },
\end{equation}
where $\theta$ is the polar angle of one lepton in the virtual photon  
rest frame and $\phi$ is the angle between the direction of
polarization and the normal to the dilepton decay plane.
The asymmetry vanishes if one integrates over the azimuthal angle. 
An estimate given in ref.~\cite{Martin} yields an 
asymmetry of the order of only one percent. 
One should stress, however, that the asymmetry level strongly 
depends on the structure function values, which
are unknown at present. Although in the non-relativistic quark model 
the relation $\delta q(x) = \Delta q(x)$ holds, in reality differences
between both distributions are expected to be caused by dynamical effects.
For the quoted estimate the somewhat arbitrary relation 
$\delta q(x,Q^2=4~GeV^2) = \Delta q(x,Q^2=4~GeV^2)$ was assumed 
and the $Q^2$-evolution equation for $\delta q(x,Q^2)$ was used
to calculate the function at other values of $Q^2$.

We note that there exist another potentially interesting 
possibility, the study of the {\bf longitudinal-transverse}
double spin asymmetry, $A^{DY}_{LT}$. 
This asymmetry was calculated in
ref.'s~\cite{JaffeJi2,Tangerman} and depends in a rather complicated fashion
on both twist-2 ($\Delta
q(x)$ and $\delta q(x)$) and twist-3 ($g_T(x)$ and $h_L(x)$) polarized
structure functions. In contrast to $A^{DY}_{LL}$ and $A^{DY}_{TT}$
the asymmetry $A^{DY}_{LT}$ decreases as $M/Q$. \\
The HERA-$\vec N$ statistical reach has not been calculated yet
for neither $A^{DY}_{TT}$ nor $A^{DY}_{LT}$.

\section{Conclusions}
%

The physics scope for the measurement of polarized nucleon-nucleon 
collisions originating 
from an internal target in the 820 GeV HERA proton beam has been summarized.
Single spin asymmetries, accessible already with the existing unpolarized 
beam, are found to be an almost unique and powerful tool to study the nature
and physical origin of twist-3 effects; even more so when taken 
in conjunction with results of other experiments. 
When measuring the polarized gluon distribution through double spin 
asymmetries in {\it photon (plus jet)} and {\it $J/\psi$ (plus jet)} 
production -- requiring a polarized HERA proton beam -- the projected 
statistical accuracies are found to be comparable to those predicted for the 
spin physics program at RHIC. 
Although both measurements explore the same $x_{gluon}$ range they are
complementary due to the different $p_T$ ranges accessible.
A measurement of
Drell-Yan pair production with both beam and target longitudinally
polarized can improve our knowledge on the polarized light sea quark 
distributions. A study of the double transverse 
and/or longitudinal-transverse Drell-Yan spin asymmetries might open
first access to the quark transversity distribution.
In addition, there is a potential to obtain significant results on
the long-standing unexplained spin asymmetries in elastic scattering.

\section*{Acknowledgements}
%
We are indebted to M.~Anselmino, O.~Teryaev and A.~Tkabladze
for very valuable comments. We thank T.~Gehrmann for having us
supplied with Drell-Yan predictions for HERA-$\vec N$.

\newpage

%

%
%

\begin{thebibliography}{99}

%
%
\bibitem{goe1}
M. G\"ockeler, Proc. of the
{\it Workshop on the Prospects of Spin Physics at HERA},
Zeuthen, 1995, DESY 95--200, ed. by J.Bl\"umlein and
W.--D. Nowak, p.339.
\vspace{-0.3em}

\bibitem{emc1}
EMC, J. Ashman et al., {\it Phys. Lett. B206}, 364 (1988)  \\
EMC, J. Ashman et al., {\it Nucl. Phys. B328}, 1 (1989)
\vspace{-0.3em}

\bibitem{Ball}
R.~D.~Ball, S.~Forte, G.~Ridolfi, {\it Nucl. Phys. B444}, 287 (1995);
\\
{\it Phys. Lett. B378}, 255 (1996)
\vspace{-0.3em}

\bibitem{142a}
E142 Coll., P. L. Anthony et al., {\it Phys. Rev. Lett. 71}, 959
(1993)
\vspace{-0.3em}

\bibitem{143a}
E143 Coll., K. Abe et al., {\it Phys. Rev. Lett. 74}, 346 (1995)  \\
E143 Coll., K. Abe et al., {\it Phys. Rev. Lett. 75}, 25 (1995)
\vspace{-0.3em}

\bibitem{smc1}
SMC, D. Adams et al., {\it Phys. Lett. B329}, 399 (1994)
\vspace{-0.3em}

\bibitem{154a}
E154 Coll., R. Arnold et al., {\it SLAC proposal E154}, 1993
\vspace{-0.3em}

\bibitem{her1}
HERMES Coll., P. Green et al., HERMES Technical Design Report,
{\it DESY-PRC 93/06}, July 1993
\vspace{-0.3em}

\bibitem{155a}
E155 Coll., R. Arnold et al., {\it SLAC proposal E155}, 1993
\vspace{-0.3em}

\bibitem{smcg2}
SMC, D. Adams et al., {\it Phys. Lett. B336}, 125 (1994)
\vspace{-0.3em}

\bibitem{143g2}
E143 Coll., K. Abe et al., {\it Phys. Rev. Lett. 76}, 587 (1996)
\vspace{-0.3em}

\bibitem{sha1}
A. Sch\"afer et al., {\it Phys. Lett. B321}, 121 (1994)
\vspace{-0.3em}

\bibitem{smc2}
SMC, B. Adeva et al., {\it Phys. Lett. B369}, 93 (1996)
\vspace{-0.3em}

\bibitem{bun1}
G. Bunce et al., {\it Particle World 3}, 1 (1992)
\vspace{-0.3em}

\bibitem{desy96-095}
W.-D. Nowak, DESY 96-095,
hep-ph/9605411, to be publ. in the Proc. of the Adriatico Research
Conference {\it Trends in Collider Spin Physics}, ICTP Trieste, 1995.
\vspace{-0.3em}

\bibitem{RSC}
RSC Coll., {\it Proposal on Spin Physics using the RHIC
Polarized Collider}, August 1992.
\vspace{-0.3em}

\bibitem{ste1}
E. Steffens, K. Zapfe--D\"uren, same Proc. as in ref.~\cite{goe1},
p.57.
\vspace{-0.3em}

\bibitem{HERAb}
HERA--B Coll., E. Hartouni et al., {\it HERA--B Design Report}, 
DESY-PRC 95/01, January 1995
\vspace{-0.3em}

\bibitem{desy96-04}
M. Anselmino et al., On Possible Future Polarized Nucleon--Nucleon
Collisions at HERA, Internal Report, {\it DESY--Zeuthen 96--04}, May
1996.
\vspace{-0.3em}

\bibitem{art}
X. Artru, J. Czyzewski and H. Yabuki, preprint LYCEN/9423 and TPJU 12/94, 
May 1994, hep-ph/9405426.
\vspace{-0.3em}

\bibitem{ans}
M. Anselmino, M. Boglione, F. Murgia, {\it Phys. Lett. B362}, 164
(1995).
\vspace{-0.3em}

\bibitem{704pi}
E704 Coll., D.L. Adams et al., {\it Phys. Lett. B264}, 462 (1991).
\vspace{-0.3em}

\bibitem{czyz}
J.~Czyzewski, {\it Acta Phys. Polon. 27}, 1759 (1996); hep-ph/9606390.
\vspace{-0.3em}

\bibitem{Phot704}
E704 Coll., D.L. Adams et al., {\it Phys. Lett. B345}, 569 (1995).
\vspace{-0.3em}

\bibitem{hammon} 
N.~Hammon, O.~Teryaev, A.~Sch\"afer, to appear in {\it Phys. Lett. B},
hep-ph/9611359
\vspace{-0.3em}

\bibitem{psicosi}
D.~Kazakov et al., Proc. of the {\it 2nd Meeting on Possible
Measurements of Singly
Polarized $p\vec{p}$ and $p\vec{n}$ Collisions at HERA}, Zeuthen,
1995, DESY Zeuthen Internal Report 95-05, ed. by H. B\"ottcher 
and W.--D. Nowak, p.43.
\vspace{-0.3em}

\bibitem{wdn-dubna-96}
W.--D.~Nowak, Proc. of the {\it 3rd Meeting on the Prospects of 
Nucleon-Nucleon Spin Physics at HERA}, Dubna, 1996, 
JINR preprint E2-96-40, DESY-Zeuthen Internal Report 96-09,
ed. by W.--D.~Nowak and O.V.~Selyugin, p.153.
\vspace{-0.3em}

\bibitem{golosk} 
S.V.Goloskokov, O.V. Selyugin, {\it  Phys. Atom. Nucl. 58}, 1894 (1995).
\vspace{-0.3em}

\bibitem{contog}
A.P.~Contogouris, B.~Kamal, O.~Korakianitis, F.~Lebessis,
Z.~Merebashvili, {\it Nucl. Phys. Proc. Suppl. 39} BC, 98 (1995)
\vspace{-0.3em}

\bibitem{gor1}
L.E. Gordon, W. Vogelsang, preprint ANL-HEP-PR-96-15/RAL-TR-96-057,
1996.
\vspace{-0.3em}

\bibitem{TT}
O.~Teryaev, A.~Tkabladze, JINR preprint E2-96-431 (1996), hep-ph/9612301
\vspace{-0.3em}

\bibitem{GehrStir}
T.K. Gehrmann, W.J. Stirling, {\it Z.Phys. C65}, 461 (1994).
\vspace{-0.3em}

\bibitem{KNT-Dubna}
V.A.~Korotkov, W.-D.~Nowak, A.~Tkabladze, 
same Proc. as in ref.~\cite{wdn-dubna-96}, p.37.
\vspace{-0.3em}

\bibitem{96-128}
M.~Anselmino et al., DESY 96-128, hep-ph/9608393,
Proc. of the {\it Workshop
on `Future Physics at HERA'}, Hamburg, 1996, ed. by G.~Ingelman,
A.De~Roeck, R.~Klanner, p.837.
\vspace{-0.3em}

\bibitem{yok1}
A.~Yokosawa, ANL-HEP-CP-96-22,
to be publ. in same Proc. as in ref.~\cite{desy96-095}
\vspace{-0.3em}

\bibitem{DYGehrStir}
T.~Gehrmann, W.J.~Stirling, same Proc. as in ref.~\cite{96-128},
p.847.
\vspace{-0.3em}

\bibitem{JaffeJi1}
R.~L.~Jaffe, X.~Ji, {\it Phys. Rev. Lett. 71}, 2547 (1993).
\vspace{-0.3em}

\bibitem{JaffeJi2}
R.~L.~Jaffe, X.~Ji, {\it Nucl.Phys. B375}, 527 (1992).
\vspace{-0.3em}

\bibitem{Martin}
O.~Martin, A.~Sch\"afer, same Proc. as in ref.~\cite{wdn-dubna-96},
p.57.
\vspace{-0.3em}

\bibitem{Tangerman}
R.~Tangerman, same Proc. as in ref.~\cite{wdn-dubna-96}, p.65.
\vspace{-0.3em}

\end{thebibliography}
\end{document}